\documentclass[final]{siamart171218}
\usepackage{srcltx}
\usepackage{times,amsbsy,amsmath,amssymb,amscd,amsfonts,mathrsfs}
\usepackage{graphicx,epstopdf}
\usepackage[hang]{subfigure}
\usepackage{threeparttable}
\usepackage{multirow}
\usepackage{dcolumn}
\usepackage{booktabs}
\usepackage{tikz}
\usetikzlibrary{arrows, snakes, backgrounds}

\usepackage{color}
\usepackage{macros}
\usepackage{algorithm}
\usepackage{algorithmic}

\newcommand{\pd}[2]{\dfrac{\partial #1}{\partial #2}}
\newcommand{\dd}{\,{\rm d}}

\def\ie{\mathrm{e}}
\newtheorem{remark}{Remark}[section]
\newtheorem{example}[theorem]{Example}
\usepackage{lineno,hyperref}
\modulolinenumbers[5]

\newcommand{\TheTitle}{A Unified Gas-kinetic Particle Method for Multiscale Photon Transport}

\title{{\TheTitle}\thanks{Submitted to the editors October 14, 2018.
\funding{The current research is supported by Hong Kong research grant council
(16207715, 16206617),  National Science Foundation of China
(11772281,91530319,11771035), and NSAF U1530401.
}}}
\author{
Weiming Li\thanks{Applied and Computational Mathematics
Division, Beijing Computational Science Research Center, Beijing 100193, China
and Department of Mathematics, Hong Kong University of
Science and Technology, Clear Water Bay, Kowloon, Hong Kong, China.
(\email{liweiming@csrc.ac.cn}).}
\and
Chang Liu\thanks{Department of Mathematics, Hong Kong University of
Science and Technology, Clear Water Bay, Kowloon, Hong Kong, China.
(\email{cliuaa@connect.ust.hk}).}
\and
Yajun Zhu\thanks{National Key Laboratory of Science and Technology on
  Aerodynamic Design and Research, Northwestern Polytechnical University,
  Xi'an, Shaanxi 710072, China.
(\email{zhuyajun@mail.nwpu.edu.cn}).}
\and
Jiwei Zhang\thanks{Applied and Computational Mathematics
Division, Beijing Computational Science Research Center, Beijing
100193, China.
(\email{jwzhang@csrc.ac.cn}).}
\and
Kun Xu\thanks{Corresponding author. Department of Mathematics, Hong Kong University of
Science and Technology, Clear Water Bay, Kowloon, Hong Kong, China and
Department of Mechanical and Aerospace Engineering, Hong
Kong University of Science and Technology, Clear Water Bay, Kowloon,
Hong Kong, China.
(\email{makxu@ust.hk}).}}

\begin{document}

\maketitle

\begin{abstract}
  In this work, we present a unified gas-kinetic particle (UGKP)
  method for the simulation of multiscale photon transport. The
  multiscale nature of the particle method mainly comes from the
  recovery of the time evolution flux function in the unified
  gas-kinetic scheme (UGKS) through a coupled dynamic process of particle
  transport and collision. This practice improves the
  original operator splitting approach in the Monte Carlo method, such
  as the separated treatment of particle transport and collision. As a
  result, with the variation of the ratio between numerical time step
  and local photon's collision time, different transport physics can
  be fully captured in a single computation. In the diffusive limit,
  the UGKP method could recover the solution of the diffusion equation
  with the cell size and time step being much larger than the photon's
  mean free path and the mean collision time.  In the free transport
  limit, it presents an exact particle tracking process as the
  original Monte Carlo method. In the transition regime, the weights
  of particle free transport and collision are determined by the ratio
  of local numerical time step to the photon's collision time. Several
  one-dimensional numerical examples covering all transport regimes
  from the optically thin to optically thick are computed to validate
  the accuracy and efficiency of the current scheme. In comparison
  with the $S_N$ discrete ordinate method, the UGKP method is based on particles and avoids 
  the discretization of particle velocity space, which does
  not suffer from the ray effect.
\end{abstract}


\section{Introduction}
The radiative transfer equation describes photon propagation in
the background medium and has important applications in the fields of
astrophysics \cite{davis2012}, atmospheric physics \cite{marshak20053d},
optical imaging \cite{klose2002optical} and so on. In this paper, we
focus on the gray radiative transfer equation with isotropic
scattering, which reads
\begin{equation}\label{eq:rt-3D}
  \dfrac{1}{c}\pd{I}{t} + \bsOmega \cdot \nabla I =
  \sigma_s \left(\dfrac{1}{4\pi}\int_{\bbS^2} I\dd\bsOmega - I\right) -
  \sigma_a I + G,
\end{equation}
where $I(t,\bx, \bsOmega)$ is the specific intensity which depends on
time $t$, space $\bx \in \bbR^3$, and angle $\bsOmega$, while $c$ is
the speed of light, $\sigma_s$ is the scattering coefficient,
$\sigma_a$ is the absorption coefficient, and $G$ is an internal
source of photons.

There are typically two categories of numerical methods for solving
the radiative transfer equations. The first category consists of the
deterministic methods with different ways of discretizing and modeling, such as the discrete ordinate
method \cite{hunter2013comparison, coelho2014advances, chen2015, roos2016conservation} and the
moment methods \cite{frank2006partial, carrillo2008numerical,
vikas2013radiation, alldredge2016approximating}. The second category
consists of the stochastic
approach, for example, the Monte Carlo method \cite{fleck1971,
lucy1999computing, hayakawa2007coupled}. The Monte Carlo method
is a very popular method for solving the radiative transfer
problems. In comparison with the deterministic methods, it is more
efficient in optically thin regime especially for the multidimensional
cases, and it does not
suffer from the ray effect. However, it has difficulties when it comes
to diffusive regime. In diffusive regime where the mean free path
is small, photons may go through a huge number of scatterings during their
lifetimes. Direct simulation of each scattering process for all
particles makes the Monte Carlo method very expensive in the diffusive regime.

On the other hand, in the diffusive regime the photon transport
process could be well described by the diffusion equation, which could
be solved efficiently. Based on this observation, many hybrid methods
have been developed in order to improve the overall efficiency in
different regimes \cite{fleck1984random, giorla1987random,
densmore2007hybrid, densmore2012hybrid}, where the Monte Carlo method
is used in the optically thin regions and the diffusion equation is
applied to the optically thick regions.  However, as far as we know,
there is still no unifying principle for accurate domain decomposition
for different regimes.

Another approach towards releasing the stiffness issue in the
diffusive regime is to develop asymptotic-preserving (AP) schemes
\cite{klar1998asymptotic, naldi1998numerical, jin1999efficient,
  jin2000uniformly, mieussens2013asymptotic, sun2015asymptotic1,
  sun2015asymptotic, sun2017multidimensional, sun2017implicit,
sun2018asymptotic} .  One of the examples is the unified gas-kinetic
scheme (UGKS), which couples the particles' transport and collision
process using a multiscale flux function obtained from the integral
solution of the kinetic model equation. The cell size and time step
are not restricted by the mean free path and mean collision time. It
was developed initially in the field of rarefied gas dynamics
\cite{xu2010unified, xu2015direct} and has been applied to the field
of radiative transfer \cite{mieussens2013asymptotic,
  sun2015asymptotic1, sun2015asymptotic, sun2017multidimensional,
  sun2017implicit, sun2018asymptotic}, plasma transport
  \cite{liu2017unified} and disperse multi-phase flow
  \cite{liu2018unified}. Since it is a discrete ordinate based
  numerical scheme, it has no statistical noise, but unavoidably
  suffers from the ray effect.

In this work, we combine the advantages of the UGKS and the Monte
Carlo method, and develop a novel unified gas-kinetic particle (UGKP) method to
describe the multiscale photon transport. In our method,
the photons are described by the particle transport and collision, and this process is
controlled by a multiscale transport solution in all regimes.
More specifically,
the Monte Carlo particle model is used to discretize the angular
direction of the photon's movement. Based on the particles' transport nature in
the discretize physical space, particles are categorized into three
groups.  Given a fixed time step, the freely transported particles are
accurately tracked by following the trajectories of the simulation
particles, while those particles that suffer collision within the
given time step are grouped and re-sampled according to the
macroscopic variables at the new time level.
The fluxes across a cell interface from different type particles are
taken into account for the updating of cell averaged macroscopic
variables. In such a way, the multiscale process through particles'
transport and their interaction through macroscopic hydrodynamics is recovered.
The multiscale flux function of the UGKS is precisely preserved in the particle
implementation. In the diffusive regime, the resulting algorithm would
become a standard central difference scheme for the diffusion
equation. In the optically thin regime, it gives a particle
tracking method same as  the Monte Carlo method. In the transition regime, the
ratio of the time step over particle collision time determines the
transport dynamics between the above two limits.

The rest of this paper is organized as follows. Section
\ref{sec:preliminary} briefly recalls the basic idea of the unified
gas-kinetic scheme (UGKS) for the linear transport equation. Section
\ref{sec:method} presents the UGKP method
for linear photon transport and the gray radiative transfer equations. In
Section \ref{sec:numerics}, numerical tests are presented to
demonstrate the accuracy and robustness of the particle method. The final
section is the conclusion.

\section{Review of the UGKS for the linear transport equation}
\label{sec:preliminary}
The unified gas-kinetic scheme (UGKS) was initially developed for the
problems in the field of rarefied gas dynamics \cite{xu2010unified,
xu2015direct}, and have also been successfully applied to problems in
radiative transfer under the finite volume framework \cite{mieussens2013asymptotic, sun2015asymptotic1, sun2015asymptotic,
sun2017multidimensional, sun2017implicit, sun2018asymptotic}. In this
section, we review the basic idea of the UGKS using the example of the
one-dimensional linear transport equation in a purely scattering
medium.

Consider
\begin{equation}
  \dfrac{1}{c}\pd{I}{t} + \mu \pd{I}{x} =
  \sigma \left(\dfrac{1}{2}\int_{-1}^1 I\dd\mu - I\right),
\end{equation}
which give a non-dimensional equation
\begin{equation}\label{eq:rt-nondimensional}
  \epsilon \pd{I}{t} + \mu \pd{I}{x} = \dfrac{\sigma}{\epsilon}
  \left(\dfrac12 E - I\right),
\end{equation}
where $E = \int_{-1}^1 I(\mu) \dd\mu$. We employed the same
non-dimensionalization process as \cite{mieussens2013asymptotic}.

The UGKS is based on a finite volume framework. We assume uniform mesh
for simplicity of discussion.
Define
\begin{equation}
  I^n_j = \dfrac{1}{\Delta
  x}\int_{x_{j-\frac12}}^{x_{j+\frac12}} I(t_n, x, \mu) \dd x
\end{equation}
to be the averaged specific intensity $I$ over a spatial cell,
and
\begin{equation}
  E^n_j = \dfrac{1}{\Delta
  x}\int_{x_{j-\frac12}}^{x_{j+\frac12}} E(t_n, x) \dd x
\end{equation}
to be the averaged energy density function $E$ over a spatial
cell. Under the finite volume framework, the discretizations of the
microscopic and macroscopic governing equations give
\begin{equation}\label{eq:rt-discretize-fvm}
  \dfrac{I^{n+1}_j - I^n_j}{\Delta t} + \dfrac{1}{\Delta
  x}\left(\phi_{j+\frac12} - \phi_{j-\frac12}\right) =
  \dfrac{\sigma}{\epsilon^2}\left(E^{n+1}_j -
  I^{n+1}_j\right),
\end{equation}
and
\begin{equation}\label{eq:rt-macro-fvm}
  \dfrac{E^{n+1}_j - E^n_j}{\Delta t} + \dfrac{1}{\Delta x}
  \left(\Phi_{j+\frac12} - \Phi_{j-\frac12}\right) = 0,
\end{equation}
where the microscopic and macroscopic flux terms are respectively
\begin{equation}\label{eq:flux-micro}
  \phi_{j+\frac12} = \dfrac{1}{\epsilon \Delta
  t}\int_{t_n}^{t_{n+1}} \mu I(t, x_{j+\frac12}, \mu) \dd t,
\end{equation}
and
\begin{equation}\label{eq:flux-macro}
  \Phi_{j+\frac12} = \int_{-1}^1 \phi_{j+\frac12}(\mu) \dd \mu.
\end{equation}

The key ingredient of the UGKS is the construction of the multiscale
flux function by adopting the integral solution of the kinetic model
equation \eqref{eq:rt-nondimensional}. Assuming  a
local constant $\sigma$, the integral solution of equation
\eqref{eq:rt-nondimensional} along the characteristic line gives
\begin{equation}\label{eq:integral-solution}
  \begin{split}
    I(t, x_{j+\frac12}, \mu) = &
  \ie^{-\frac{\sigma_{j+\frac12}(t - t_n)}{\epsilon^2}} I\left(t_n,
    x_{j+\frac12} -
  \frac{\mu}{\epsilon}(t - t_n)\right) \\
  & + \int_{t_n}^t
  \ie^{-\frac{\sigma_{j+\frac12}(t-s)}{\epsilon^2}}
  \times \dfrac{\sigma_{j+\frac12}}{\epsilon^2} \frac12 E\left(s, x_{j+\frac12} -
  \frac{\mu}{\epsilon}(t - s)\right) \dd s,
\end{split}
\end{equation}
which is used to construct the numerical fluxes in equation
\eqref{eq:rt-discretize-fvm}.The integral solution couples transport
with particle collisions, and bridges the kinetic and the
hydrodynamic scale dynamics.

The  numerical fluxes for microscopic and macroscopic variable updates are based on the
piecewise linear initial reconstruction of $I$ and $E$ at the beginning of each time step.
The details were presented in  \cite{mieussens2013asymptotic} and \cite{sun2015asymptotic1}.
It has been proved in \cite{mieussens2013asymptotic} that when $\sigma$ equals $0$, the
UGKS tends to the finite volume scheme
\begin{equation}
  \dfrac{I^{n+1}_j - I^n_j}{\Delta t} + \dfrac{1}{\Delta
  x}\dfrac{\mu}{\epsilon} \left(\left(I^n_j 1_{\mu >
    0} + I^n_{j+1} 1_{\mu < 0}\right) - \left(I^n_{j-1} 1_{\mu>0} +
  I^n_j 1_{\mu < 0}\right)\right) = 0,
\end{equation}
which is consistent with free transport solution.
In the diffusion limit, with a uniform mesh the UGKS
scheme becomes
\begin{equation}
  \dfrac{E^{n+1}_j - E^n_j}{\Delta t} - \dfrac{1}{\Delta
  x}\left(\dfrac{1}{3 \sigma_{j+\frac12}}\dfrac{E^{n+1}_{j+1} -
    E^{n+1}_j}{\Delta x} - \dfrac{1}{3
    \sigma_{j-\frac12}}\dfrac{E^{n+1}_j -
  E^{n+1}_{j-1}}{\Delta x}\right) = 0,
\end{equation}
which is a standard central difference scheme for
the limit diffusion equation as $\epsilon$ tends to $0$. For more
details on the asymptotic analysis of the UGKS for the radiative
transfer equation we refer to \cite{mieussens2013asymptotic} and
\cite{sun2015asymptotic1}.

Following the methodology of the UGKS, we will construct a particle
algorithm with multiscale transport property for recovering transport physics from  the kinetic
scale to the hydrodynamic scale. For the kinetic scale particle free transport, the method tracks the particle trajectory
precisely; for those particles suffering collisions, the updated macroscopic variables will be used
to re-sample them.
A multiscale particle method for equations \eqref{eq:flux-macro},
\eqref{eq:flux-micro}, and \eqref{eq:integral-solution}
is constructed through the tracking and re-sampling particles with the help of updated macroscopic variables.

\section{Multiscale Particle Method}
\label{sec:method}
In this section, we will first show the physical picture for the
particle classification and evolution. Then, the multiscale particle algorithm will be introduced.
This algorithm is first presented for a single linear transport equation;
then it will be extended to the standard one-group radiative transfer and material temperature
equations.

\subsection{Classification of particles}\label{sec:classification-particle}
\begin{figure}[htbp]
  \centering
  \begin{tikzpicture}[scale=0.5]
  \draw (-2, -4)--(-2, 5);
  \draw (-7, 3) circle(8pt);
  \draw[->, thick] (-6.5,3) -- (2.5,3);
  \draw (-0.5, 3.5) node {free transport};
  \draw (3, 3) circle(8pt);

  \draw (-9, 1) circle(8pt);
  \draw[->, thick] (-9.1,0.6) -- (-9.9,-2.6);
  \draw (-11.5, -1.5) node {free transport};
  \draw (-10, -3) circle(8pt);

  \filldraw[gray] (-4, 0) circle(8pt);
  \draw[->, thick] (-3.5,0) -- (4.5,0);
  \draw (-0.5, 0.5) node {transport};
  \filldraw[gray] (5,0) circle(8pt);
  \draw[dashed] (5,0) circle(20pt);
  \draw[snake] (5.5,-0.2) -- (6.7, -2.3);
  \draw[->, thick] (6.7,-2.3) -- (7, -2.5);
  \draw (7.5, -1.2) node {collides};
  \filldraw[gray] (7.3,-2.8) circle(8pt);

  \filldraw[black] (-6.7,0) circle(8pt);
  \draw[dashed] (-6.7,0) circle(20pt);
  \draw[snake] (-6.5,-0.2) -- (-3.2, -2.3);
  \draw[->, thick](-3.2,-2.3) --(-1.6,-3.1);
  \draw (-4.5, -1.2) node {collides};
  \filldraw[black] (-1.3,-3.1) circle(8pt);
  \draw (-2.5,-4.5) node {numerical cell interface};
\end{tikzpicture}
\caption{Diagram for classification of particles.}\label{fig:classification-particle}
\end{figure}
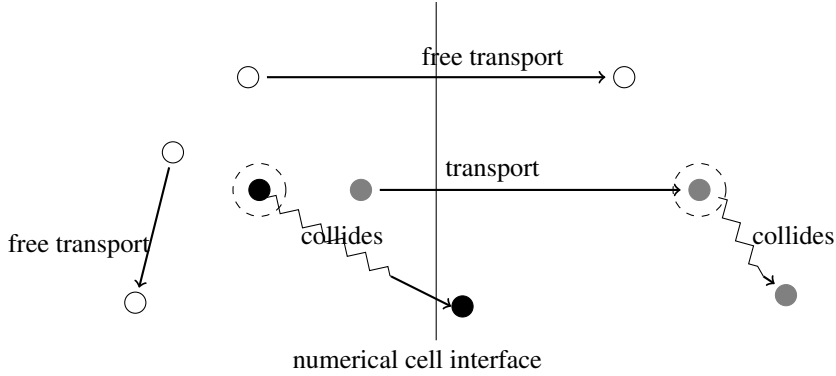

The particles can move freely until they interact
with background medium.
Based on the process of transport and collision, the particles can be divided into three types, which
are denoted as Type I, Type II, and Type III particles.
Type I particles travel freely within the entire time step.
Type II particles travel freely across a cell
interface before they collide with the background medium. Type III
particles collide with the background medium before they reach the
cell interface. Note that within a whole time step
any particle of the three types can stay
in the same cell or move to the neighboring cell.
More specifically, Type I particles may transport freely
to the neighboring cell during the time step, or they transport to another place within the same cell.
Type II particles may move freely to the neighboring cell,
collide with the background medium, and then remain in the same
cell for the rest of the time step, or they may bounce back to
their former cell or transport to another neighboring cell after collision.
Type III particles may stay in the same cell after their first collision
with the background medium until the end of the time step, or they may
move across the cell interface before the end of the time step.
This classification of particles is illustrated in Figure
\ref{fig:classification-particle}, where Type I particles are denoted
by white circles, Type II particles by grey circles, and Type III
particles by black circles.
Denoting $t_c$ as the time for the first collision event for each
particle and $t_{in}$ as the time for each particle to freely
transport until it reaches the cell interface, the
conditions for classification of particles are summarized in Table
\ref{tab:classify-particle}.
\begin{table}[htbp]
    \centering
  \caption{Classification of particles.}\label{tab:classify-particle}
\setlength{\tabcolsep}{11mm}{
    \begin{tabular}{cc}
      \toprule
      Particle type & Condition \\
        \hline
        Type I & $t_c > \Delta t$ \\
        \hline
        Type II & $ \Delta t > t_c > t_{in}$ \\
        \hline
        Type III & $t_c < \Delta t ~\&~ t_c < t_{in}$\\
        \bottomrule
      \end{tabular}}
\end{table}

Note that the UGKS solves the linear kinetic equation by a finite
volume method and the flux terms are approximated by
Eq.{\eqref{eq:integral-solution}}. In order to recover the multiscale transport in Eq.{\eqref{eq:integral-solution}},
the free transport process of Type I and
Type II particles mimics the first term in
Eq.\eqref{eq:integral-solution}, while the collision effect of Type II
and Type III particles simulate the second term in
\eqref{eq:integral-solution}. In this way, we can recover the UGKS flux
through particle implementation.  Type I particles and the free
transport process of Type II particles can be tracked precisely, while
Type II particles after collision and Type III particles are grouped
and are re-sampled from macroscopic variables due to their close connection to the equilibrium state.
The detailed implementation is given in Section
\ref{sec:linear-transport-algorithm}.

\subsection{The unified gas-kinetic particle method}\label{sec:linear-transport-algorithm}
Under the Monte Carlo framework, the
specific intensity $I(t, x, \mu)$ is represented by a finite number of simulation
particles. For each particle, the unknown variables are its position,
velocity, and weight, denoted as the 3-tuple $(x_j, \mu_j, w_j)$. The
computation domain is divided into cells to locate
particles and sample the local macroscopic quantities. Denote $E_m$ as
the average of the macroscopic energy density within cell $m$, $V_m$ as
the volume of cell $m$,  and $w_j$ as the weight of the $j$-th
particle within the same cell. Denote $N_m$ as the number of simulation
particles within cell $m$. Then, the macroscopic quantities and the
corresponding particle information satisfy the following
relationship:
\begin{equation}
  E_m = \dfrac{1}{V_m}\sum\limits_{j = 1}^{N_m} w_j.
\end{equation}

This section considers the particle method for solving equation \eqref{eq:rt-nondimensional}, which is
the linear equation with purely scattering medium. The extension of the method to
the radiation-material coupled equations  will be discussed
in Section \ref{sec:coupled}.

With the approximation of $\sigma$ as a local constant value $\sigma_j$ and using an implicit approximation
to $E$, Eq. \eqref{eq:integral-solution} can be modeled as
\begin{equation}\label{eq:time-stepping-characteristic}
  I\left(t_{n+1}, x\right) =
  \ie^{-\frac{\sigma}{\epsilon^2} \Delta t} I\left(t_n,
    x-\frac{\mu}{\epsilon}\Delta t\right) + \left(1 -
  \ie^{-\frac{\sigma}{\epsilon^2} \Delta t} \right) \frac12
  E\left(t_{n+1}, x\right).
\end{equation}
Eq. \eqref{eq:time-stepping-characteristic} could be implemented
under the particle Monte Carlo framework.
For a fixed time step $\Delta t$, each
particle is allowed to transport freely under probability
$\ie^{-\frac{\sigma}{\epsilon^2} \Delta t}$.
At the same time, with the probability $1 -
\ie^{-\frac{\sigma}{\epsilon^2} \Delta t}$, the particle is
re-sampled from the equilibrium distribution at the new time step.
The algorithm consists of the following
steps: the free transportation of particle is precisely followed and
contributes to the kinetic scale fluxes. Together with the
hydrodynamic scale fluxes obtained from the equilibrium state,
the macroscopic variables can be updated first, then the updated
macroscopic quantities guide the updating of the microscopic
particle distribution. In the above procedures, both macroscopic and
microscopic quantities will be updated within each control volume.
In the next section, the  details for updating macroscopic variable will be discussed.

\subsubsection{Updating macroscopic quantity}
To simplify discussions, the method will be given under the assumption
of the uniform mesh. Its extension to non-uniform mesh is
straightforward. Eq. \eqref{eq:rt-macro-fvm}
for updating the
macroscopic variable can be re-written as
\begin{equation} \label{eq:macro-update}
  E_{j}^{n+1} = E_j^n - \dfrac{\Delta t}{\Delta x}
  \left(\Phi_{j+\frac12} - \Phi_{j-\frac12}\right).
\end{equation}
The interface fluxes \eqref{eq:flux-macro} are based on the solution of Eq. \eqref{eq:integral-solution},
\begin{equation}\label{eq:macro-flux-integral}
  \begin{split}
    \Phi_{j+\frac12} = & \dfrac{1}{\epsilon \Delta
  t}\int_{t_n}^{t_{n+1}}\int_{-1}^1 \mu
  \ie^{-\frac{\sigma_{j+\frac12}(t-t_n)}{\epsilon^2}} I\left(t_n,
  x_{j+\frac12}-\frac{\mu}{\epsilon}(t-t_n), \mu\right)\dd\mu \dd t \\
  & + \frac{1}{2\epsilon \Delta t}
  \int_{t_n}^{t_{n+1}}\int_{t_n}^t\int_{-1}^1 \mu
  \frac{\sigma_{j+\frac12}}{\epsilon^2}
    \ie^{-\frac{\sigma_{j+\frac12}(t-s)}{\epsilon^2}}
    E\left(s,x_{j+\frac12}-\frac{\mu}{\epsilon}(t-s)\right)\dd s \dd\mu \dd t.
  \end{split}
\end{equation}
With   the  piecewise linear reconstruction for $E$,
\begin{equation}
  \begin{split}
    E\left(s, x_{j+\frac12} - \frac{\mu}{\epsilon} (t-s)\right) = & E(t_{n+1},
  x_{j+\frac12}) + \pd{E}{t}(t_{n+1}, x_{j+\frac12}) \times (s -
  t_{n+1}) \\
  & +
  \pd{E}{x}(t_{n+1}, x_{j+\frac12}) \times
  \left(-\frac{\mu}{\epsilon}(t - s)\right),
\end{split}
\end{equation}
 the implicit central difference discretization for $\pd{E}{x}$,
\begin{equation}
  \pd{E}{x}(t_{n+1}, x_{j+\frac12})
  \approx  \dfrac{E^{n+1}_{j+1} - E^{n+1}_j}{\Delta x},
\end{equation}
and the direct computation
\begin{equation}
  \begin{split}
    & \int_{t_n}^{t_{n+1}} \left(\int_{t_n}^t (t - s)
  \ie^{-\frac{\sigma(t-s)}{\epsilon^2}} \dd s \right) \dd t \\
  = & \frac{1}{\left(\frac{\sigma}{\epsilon^2}\right)^2}\left(-\dfrac{2}{\left(\frac{\sigma}{\epsilon^2}\right)}\left(1
    - \ie^{-\frac{\sigma \Delta t}{\epsilon^2}} \right) + \Delta t
  \left(1 + \ie^{-\frac{\sigma \Delta t}{\epsilon^2}}\right)\right),
\end{split}
\end{equation}
the update $E^{n+1}$ in Eq. \eqref{eq:macro-update} becomes
\begin{equation}\label{eq:macro-eq}
  \dfrac{\Delta t}{\Delta x^2} \alpha_{j-\frac12} E^{n+1}_{j-1} +
  \left(1 - \dfrac{\Delta t}{\Delta x^2}\left(\alpha_{j-\frac12} +
  \alpha_{j+\frac12}\right)\right) E^{n+1}_j + \dfrac{\Delta t}{\Delta
  x^2}\alpha_{j+\frac12} E^{n+1}_{j+1} = r.h.s,
\end{equation}
where
\begin{equation}
  \alpha_{j+\frac12} = -\frac{1}{3 \sigma_{j+\frac12}} \left(-\dfrac{2}{\left(\frac{\sigma_{j+\frac12} \Delta t}{\epsilon^2}\right)}\left(1
    - \ie^{-\frac{\sigma_{j+\frac12} \Delta t}{\epsilon^2}} \right) +
  \left(1 + \ie^{-\frac{\sigma_{j+\frac12} \Delta
t}{\epsilon^2}}\right)\right),
\end{equation}
and
\begin{equation}\label{eq:macro-flux-free-transport}
  \begin{split}
    r.h.s = E^n_j & + \dfrac{1}{\Delta x}\int_{-1}^1
    \int_{t_n}^{t_{n+1}} \dfrac{\mu}{\epsilon}
    \ie^{-\frac{\sigma_{j-\frac12}(t-t_n)}{\epsilon^2}}
  I\left(t_n, x_{j-\frac12} - \frac{\mu}{\epsilon} (t - t_n),
    \mu\right) \dd t
  \dd\mu \\
  & - \dfrac{1}{\Delta x}\int_{-1}^1 \int_{t_n}^{t_{n+1}}
  \dfrac{\mu}{\epsilon} \ie^{-\frac{\sigma_{j+\frac12}(t - t_n)}{\epsilon^2}}
  I\left(t_n, x_{j+\frac12} - \frac{\mu}{\epsilon} (t - t_n),
    \mu\right) \dd t
  \dd\mu.
\end{split}
\end{equation}
The Monte Carlo implementation of the right hand side of Eq.
\eqref{eq:macro-flux-free-transport} for the computation of the fluxes
is about calculating the freely transported particles across the cell
interface, and the number of particles across the interface is
computed during the transport process while taking into account the
possible particle collisions.  After determining its right hand side,
Eq. \eqref{eq:macro-eq} can be solved to get $E^{n+1}$.  Subsequently,
it can be used to re-sample particles.

\subsubsection{Updating particle distribution}\label{sec:update-particle}
For updating particle distribution, the free transport process of Type
I and Type II particles before collision with background medium are
tracked precisely. The influence of the collision process on the particle distribution (Type
II and Type III particles) is considered by re-sampling according to
the equilibrium distribution at the new time step.
We denote $W$ to be the total energy density of
Type II and Type III particles. Our algorithm for updating particle
information within each time step is as follows:
\begin{enumerate}
  \item At the beginning of a time step, set the macroscopic variable $W$
    to zero.
  \item Perform the following for all particles: for each particle,
    generate time $t_c$ at which the first collision event happens
    according to the local $\sigma$. There are three possible
    scenarios:
    \begin{enumerate}
      \item If $t_c > \Delta t$, the particle is allowed to transport
        freely.
      \item If $\Delta t > t_c > t_{in}$, the weight of the particle
        is added to $W$ of the neighbouring cell where the particle goes.
      \item If $t_c < t_{in}$ and $t_c < \Delta t$, the weight of the
        particle is added to $W$
        of the current cell.
    \end{enumerate}
  \item Add the contribution to $W$ in the cells through the macroscopic flux
    $\Phi_{j+\frac12}$ denoted as $\Phi_{E}$,
    \begin{equation}\label{eq:macro-flux-equilibrium}
   \Phi_{E,j+\frac12} = \alpha_{j+\frac12}
  \dfrac{E^{n+1}_{j+1} - E^{n+1}_j}{\Delta x}.
\end{equation}
  \item Generate particles according to the equilibrium distribution constructed from
    $W$. Specifically, a piecewise linear reconstruction
    of $W$ is constructed first. Then, the position of the particles are sampled according to
    the distribution of $W$ in space, while the microscopic velocities of the
    particles are sampled according to a uniform distribution on
    $[-1,1]$.
\end{enumerate}
\begin{remark}
  Due to the stochastic noise and high order reconstruction of $W$, when $W$ is very close to zero, its
  numerical values could sometimes be negative. This could be treated
  either by setting $W$ to zero, or by generating particles with
  negative weights.
\end{remark}

\subsubsection{Outline of the algorithm}
In this section, we give a brief summary of the implementation of the
unified gas-kinetic particle method.

Upon initialization, we use a weight factor $w_p$ to indicate the
amount of photon energy represented by each simulation particle.  Given
the initial condition of $E$ in cell $m$, one can obtain the number of
particles $N_m$ by
\begin{equation}
  N_m = \dfrac{E_m V_m}{w_p}.
\end{equation}
Then, the local initial distribution function $I(t_0, x_j, \mu)$ is applied to
generate $N_m$ particles and their initial velocities. Given the
information of particles and the macroscopic energy density,
the system would then be evolved to the next time step by Algorithm
\ref{alg:flowchart}. This is done for each time step until the end of
computation.
\begin{algorithm}[htbp]
\caption{ The unified gas-kinetic particle method within one time step
$\Delta t$.}
\label{alg:flowchart}
\begin{algorithmic}[1]
\REQUIRE ~~\\
The set of $(x_i, \mu_i, w_i)$ for all particles at time $t = t_n$;\\
The set of macroscopic variables $E_j$ for all spatial cells at time
$t = t_n$;\\
\ENSURE ~~\\
The set of $(x_i, \mu_i, w_i)$ for all particles at time $t = t_{n+1}$;\\
The set of macroscopic variables $E_j$ for all spatial cells at time
$t = t_{n+1}$;\\
\STATE Set the macroscopic variable for re-sampling. $W$ is $0$ for all
spatial cells and $N_{total} $ is the total current number of simulation particles;
\label{ code:flow:initialize }
\FOR{ each $i \leq N_{total}$}
\STATE generate the time at which the first collision happens: $t_c =
-\dfrac{\ln \eta}{\frac{\sigma}{\epsilon^2}}$, where $\eta\in(0,1)$ is a random
number from uniform distribution;
\STATE find the time for free transport $t_f = \min(t_c, \Delta t)$;
\STATE update particle position $x_i = x_i + \dfrac{\mu_i}{\epsilon} t_f$;
\IF{ $t_c \leq \Delta t$}
\STATE find $j=$ index of cell for $x_i$;
\STATE add the weight of this particle to macroscopic variable for
re-sampling: $W_j = W_j + \dfrac{w_i}{\Delta x}$;
\STATE delete information of this particle $(x_i, \mu_i, w_i)$;
\ENDIF
\ENDFOR;
\STATE Calculate $E^{n+1}$ by solving equation \eqref{eq:macro-eq};
\STATE Calculate $\Phi_{E}$ by equation \eqref{eq:macro-flux-equilibrium};
\STATE Evolve $W$ by $\Phi_E$ to $W^\ast$ using $W^\ast_j = W_j -
\dfrac{\Delta t}{\Delta x}\left(\Phi_{E,j+\frac12} -
\Phi_{E,j-\frac12}\right)$; \STATE Re-sample particles from $W^\ast$
as described in Section \ref{sec:update-particle};
\RETURN Distribution of all particles $\bigcup (x_i, \mu_i, w_i)$ and $E^{n+1}$.
\end{algorithmic}
\end{algorithm}

\subsection{Properties of the algorithm}
The unified gas-kinetic particle method satisfies the following properties:
\begin{enumerate}
  \item The energy density is conserved.
  \item The macroscopic variable $E$ is the summation of contribution from all
    microscopic particles.
  \item In the diffusive limit, $\epsilon \rightarrow 0$ and
    $\ie^{-\frac{\sigma_j \Delta t}{\epsilon^2}} \rightarrow 0$, each
    particle is re-sampled from the equilibrium distribution with
    probability $1$. At the same time, the scheme for updating
    macroscopic variables tends to the following limiting equation
    \begin{equation}
      \dfrac{E^{n+1}_j - E^n_j}{\Delta t} - \dfrac{1}{\Delta
      x}\left(\dfrac{1}{3 \sigma_{j+\frac12}}\dfrac{E^{n+1}_{j+1} -
        E^{n+1}_j}{\Delta x} - \dfrac{1}{3
        \sigma_{j-\frac12}}\dfrac{E^{n+1}_j -
      E^{n+1}_{j-1}}{\Delta x}\right) = 0.
    \end{equation}
    The algorithm is equivalent to a time-implicit central difference
    solver of the diffusion equation.
  \item In the free transport limit, $\sigma \rightarrow 0$ and
    $\ie^{-\frac{\sigma_j \Delta t}{\epsilon^2}} \rightarrow 1$,
    each particle is traced exactly by free transport with
    probability 1. In this case, the algorithm could recover the
    exact solution for each particle.
\end{enumerate}

\subsection{Extension to the coupled equations of gray radiative transfer and material energy}
\label{sec:coupled}
This section extends the unified gas-kinetic
particle method to solve the coupled system of gray radiative transfer equation and material
temperature equation,
\begin{equation}\label{eq:rt-coupled}
  \left\{\begin{split}
  & \dfrac{\epsilon^2}{c} \pd{I}{t} + \epsilon\mu \pd{I}{x} = \sigma \left(\dfrac12
    a c T^4 - I\right), \\
    & \epsilon^2 C_v \pd{T}{t} = \sigma \left(\int_{-1}^1 I(\mu) \dd\mu - a c
  T^4\right).
\end{split}\right.
\end{equation}
Define $u_r = a T^4$ and $\beta = \pd{u_r}{T}$, then the second equation
could be re-written as
\begin{equation}\label{eq:energy}
  \pd{u_r}{t} = C_v^{-1} \beta \dfrac{\sigma}{\epsilon^2} \left(\int_{-1}^1 I(\mu) \dd\mu
  - c u_r\right).
\end{equation}
The implicit Monte Carlo method proposed by Fleck and Cummings in \cite{fleck1971} has been shown to be an effective technique for
solving non-linear, time-dependent, radiative transfer problems
and is widely used in the radiative transfer
community. Fleck's implicit Monte Carlo method uses an effective scattering process to approximate
the absorption and emission of radiation by the background medium.
This treatment allows it to take larger time steps than that in a purely
explicit method. Here the similar semi-implicit discretization for
material temperature will be employed.
Specifically, Eq. \eqref{eq:energy} is discretized by
\begin{equation}
  \dfrac{u^{n+1}_r - u^n_r}{\Delta t} = C_v^{-1}\beta^n
  \dfrac{\sigma}{\epsilon^2} \left(E
- c u^{n+1}_r\right),
\end{equation}
which gives
\begin{equation}\label{eq:discretize-energy}
  u^{n+1}_r = \dfrac{1}{1 + c C_v^{-1}\beta^n \times \frac{\sigma \Delta
  t}{\epsilon^2}} u^n_r +
  \dfrac{C_v^{-1} \beta^n \times \frac{\sigma \Delta t}{\epsilon^2}}{1 + c C_v^{-1} \beta^n
  \times \frac{\sigma \Delta t}{\epsilon^2}} E.
\end{equation}
With the definition
\begin{equation}
  \sigma_a = \dfrac{\sigma}{1 + c C_v^{-1} \beta^n \times
  \frac{\sigma \Delta t}{\epsilon^2} },
  \quad \sigma_s = \sigma - \sigma_a,
\end{equation}
substituting Eq. \eqref{eq:discretize-energy} into Eq. \eqref{eq:rt-coupled} yields
\begin{equation}
  \dfrac{\epsilon^2}{c} \pd{I}{t} + \epsilon\mu \pd{I}{x} = \sigma_s\left(\frac12
  E - I \right) + \sigma_a\left(\frac{1}{2} c u^n_r - I \right).
\end{equation}
An operator splitting scheme is used to solve the above system, i.e.,
for the linear kinetic equation
\begin{equation}\label{eq:split-scattering}
  \dfrac{\epsilon^2}c\pd{I}{t} + \epsilon \mu \pd{I}{x} =
  \sigma_s\left(\dfrac12 E - I\right),
\end{equation}
and the radiation energy exchange,
\begin{equation}\label{eq:split-exchange}
  \dfrac{\epsilon^2}{c} \pd{I}{t} = \sigma_a \left(\dfrac12
  c u^n_r - I\right),
\end{equation}
with the update of material energy through conservation principle.
Here Eq. \eqref{eq:split-scattering} is solved using the algorithm introduced
in Section \ref{sec:linear-transport-algorithm}.

The numerical procedure for the updates of radiation-material coupling system is
the following.
The spatial domain is covered by the mesh points.
The variables are the cell averaged  $E$ and
$u_r$, as well as the particle positions, velocities, and weights. The
discretized absorption coefficient is defined in each cell. First,
particle distribution and $E$ are advanced to the next time step by
solving Eq. \eqref{eq:split-scattering} using Algorithm
\ref{alg:flowchart}. After the determination of the particle distribution  by
solving Eq. \eqref{eq:split-exchange},
 the energy change of particles is added to update $u_r$ through the energy conservation.
 The process is repeated for the new time step until the end of computation.

\section{Numerical Experiments}
\label{sec:numerics}
In this section, we present numerical examples to validate the
proposed UGKP method. As we are targeting  to develop
a method that automatically bridges the optically thin and optically
thick regimes, the test cases cover the rarefied
($\epsilon \gg \Delta x$), the intermediate ($\epsilon \approx
\Delta x$), and the diffusive ($\epsilon \ll \Delta x$) regimes,
as defined in \cite{jin1998diffusive}.
Depending on the regimes, the
numerical results are compared with the solutions of discrete
ordinate method and those of the diffusion equation.  A large number
of grid points are used to ensure the convergence of the reference
solutions. All numerical tests are conducted in the one-dimensional slab
geometry.

For the UGKP method, the time step is determined by $\Delta t = CFL *
\epsilon \Delta x / c$, with $CFL = 0.8$. Therefore, when $\epsilon$
is small, i.e. in the diffusive regime, the current method can use a
much larger cell size and time step than the particle mean free path
and collision time.

\subsection{Linear transport equation}
Examples in this section are for the linear equation with a possible
source term
\begin{equation}
  \epsilon\pd{I}{t} + \mu \pd{I}{x} =
  \dfrac{\sigma}{\epsilon}\left(\dfrac12 E - I\right) + \epsilon G.
\end{equation}
In the following, the results of the unified gas
kinetic particle method are obtained using $200$ grids in space, and $400$ simulation
particles within each cell. The final results are from the averages of
$10$ runs.

\begin{example}[Diffusive regime]
  Take $\epsilon = 10^{-4}$ and $G = 0$. In this example, we consider
  a semi-infinite spatial domain $x \in [0, \infty)$ with an isotropic
    inflow condition imposed on the left boundary. The numerical simulation is in a spatial domain
    $[0, 1]$.  The initial value is $I(\mu) = 0$ for all $x$.
    Inflow boundary condition is imposed at $x = 0$ with the
    incoming specific intensity $I(t,0,\mu)= \frac12$.
  \end{example}

  The reference solution is obtained from solving the diffusion
  equation with implicit discretization in time and central
  differencing in space using $200$ grids. Boundary conditions for both
  the diffusion equation and the macroscopic equation in the unified
  gas-kinetic particle method are given by $E_{ghost} = 2 E_{bd} -
  E_0$, where $E_{ghost}$ and $E_{bd}$ are the values of $E$ in the
  ghost cell and the boundary cell respectively. Results for the numerical
  solution of $E$ are compared at time $t = 0.15$.

\begin{figure}[htbp]
    \centering
    \includegraphics[width=0.48\textwidth]{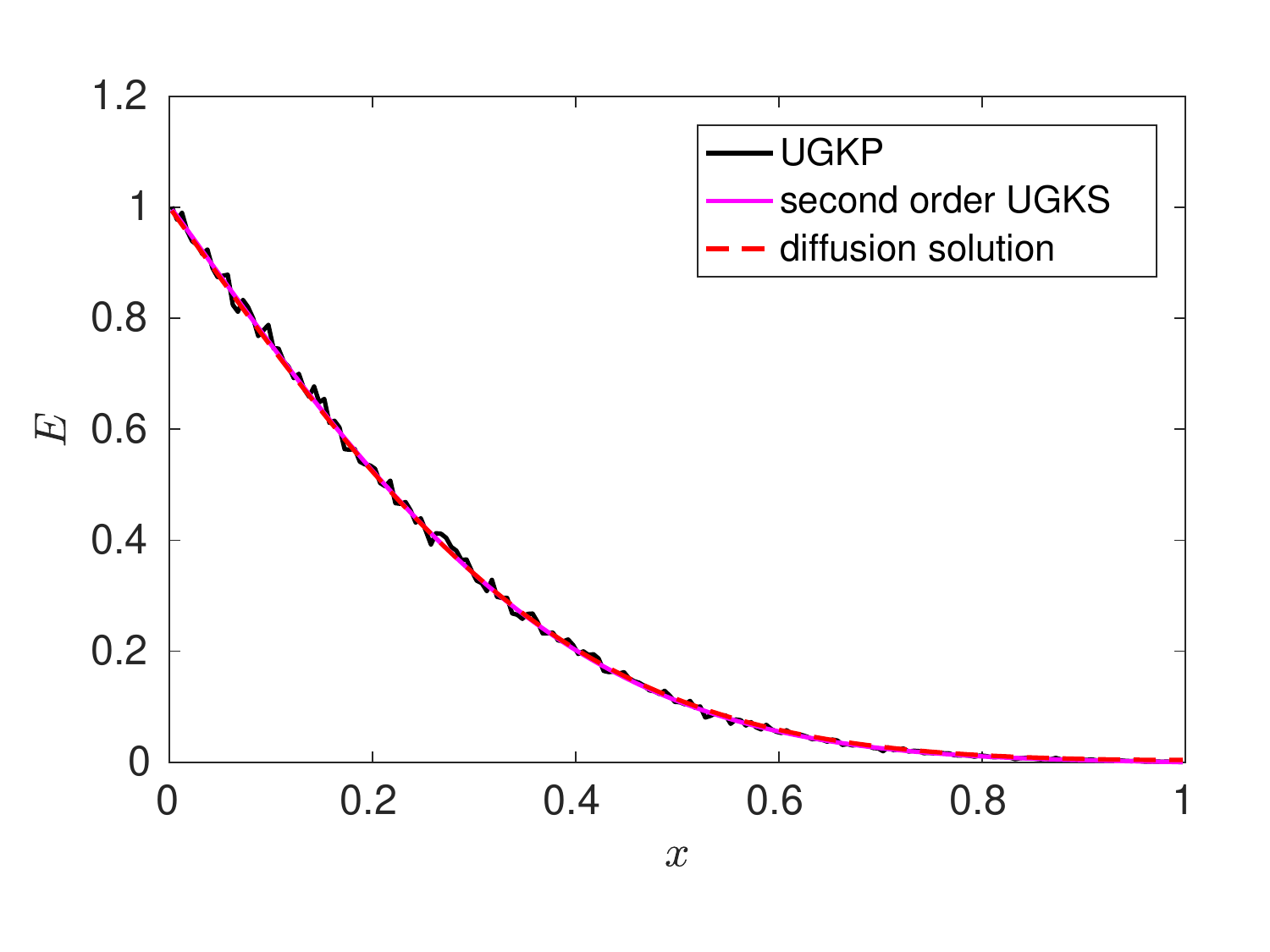}
    \caption{The macroscopic energy densities $E$ as functions of the spatial
      coordinate $x$ for the diffusive regime at $t =
    0.15$.}\label{fig:linear-diffusive}
  \end{figure}

  This example tests the UGKP method's ability
  to recover the diffusive regime. Fig. \ref{fig:linear-diffusive}
  shows the solutions from the current scheme, the second order UGKS, and the
 diffusion equation. These three solutions agree with each
  other very well. It shows that the unified
  gas-kinetic particle method can recover the diffusive solution accurately even with the mesh size being much
  larger than the photon's mean free path.

  It should be emphasized that  the size of the time step taken in our
  computation is of the order $10^{-6}$, while the mean collision time is
  of the order $10^{-8}$. Therefore, the time step used in the current scheme is around two orders of
  magnitude larger than the particle mean collision time. This advantage
  will become even more obvious for smaller $\epsilon$.

\begin{example}[Rarefied regime]\label{eg:kinetic-linear}
 The initial and boundary conditions are taken to be the same as the
 previous example, and also $G = 0$. For this test case, we take
 $\epsilon = 1$ and run the computation until $t = 0.9$.
\end{example}

\begin{figure}[htbp]
    \centering
    \includegraphics[width=0.48\textwidth]{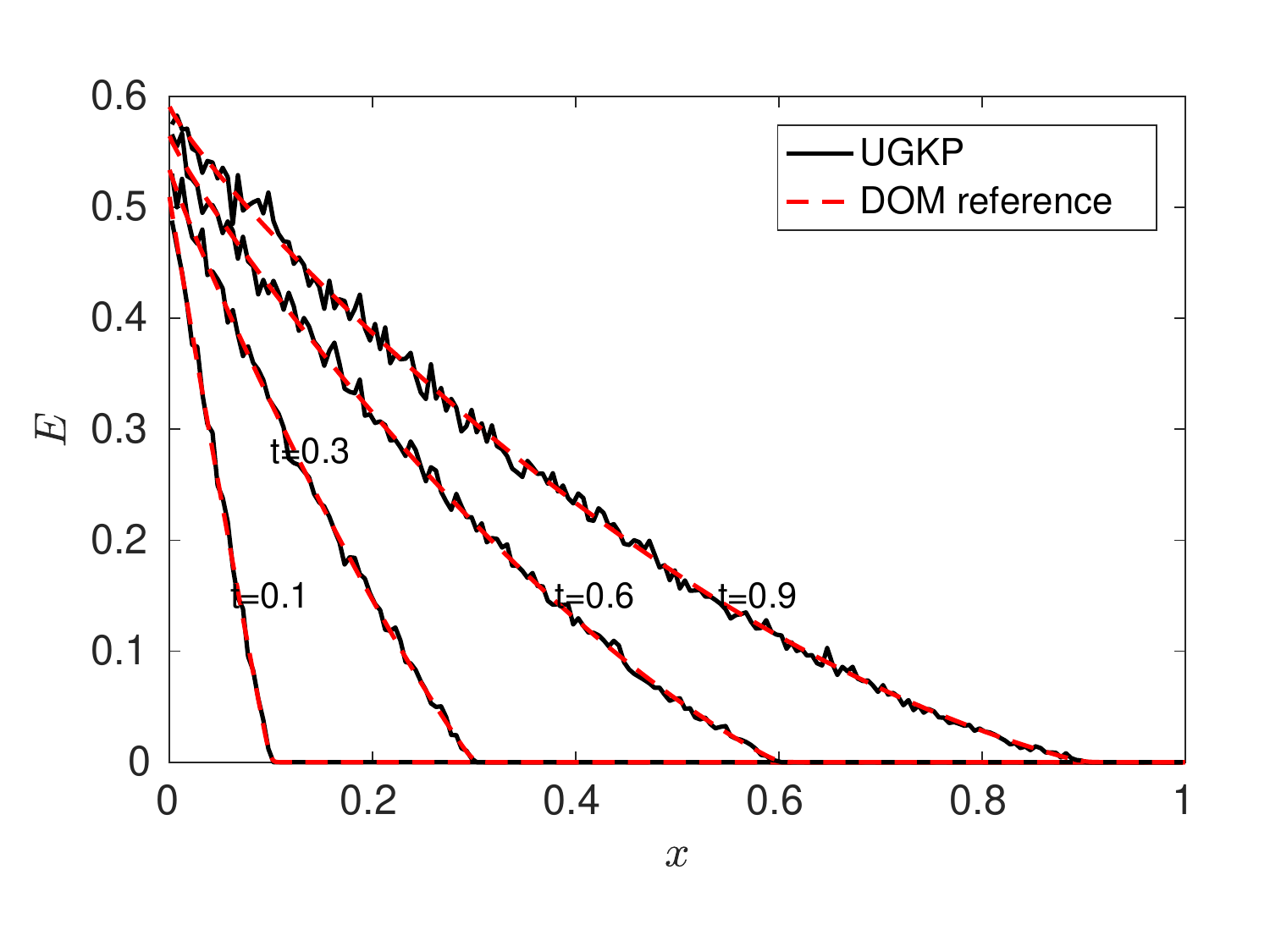}
    \caption{The macroscopic energy densities $E$ as functions of the spatial
      coordinate $x$ for the rarefied regime at $t = 0.1$,
    $0.3$, $0.6$ and $0.9$.}\label{fig:linear-kinetic}
\end{figure}

The discrete ordinates method with standard upwind discretization is
employed to get the reference solution with  $280$ points in velocity
space and $2000$ points in physical space. In Figure
\ref{fig:linear-kinetic} the results of $E$ are plotted at times $t =
0.1$, $0.3$, $0.6$ and $0.9$. It is observed that the current solutions have
excellent agreement with the reference solutions.
This shows the UGKP method
could recover accurate solution in the rarefied regime.

\begin{example}[Intermediate regime with a source term]
  In this problem, the internal radiation source is given by
  \[
    G =\left\{\begin{array}{l}
        1,~~\text{if}~~0.4\leq x\leq 0.9,\\
        0,~~\text{otherwise}.
      \end{array}\right.
    \]
  We take $\epsilon = 10^{-2}$ and $\sigma = 1 + (2x)^5$.  The computation
  domain is $x\in[0, 1]$. The initial value is set as $I = 0$ for all $x$.
  The simulation time interval is from $t_0=0$ to $t=0.02$.
\end{example}

\begin{figure}[htbp]
    \centering
    \includegraphics[width=0.48\textwidth]{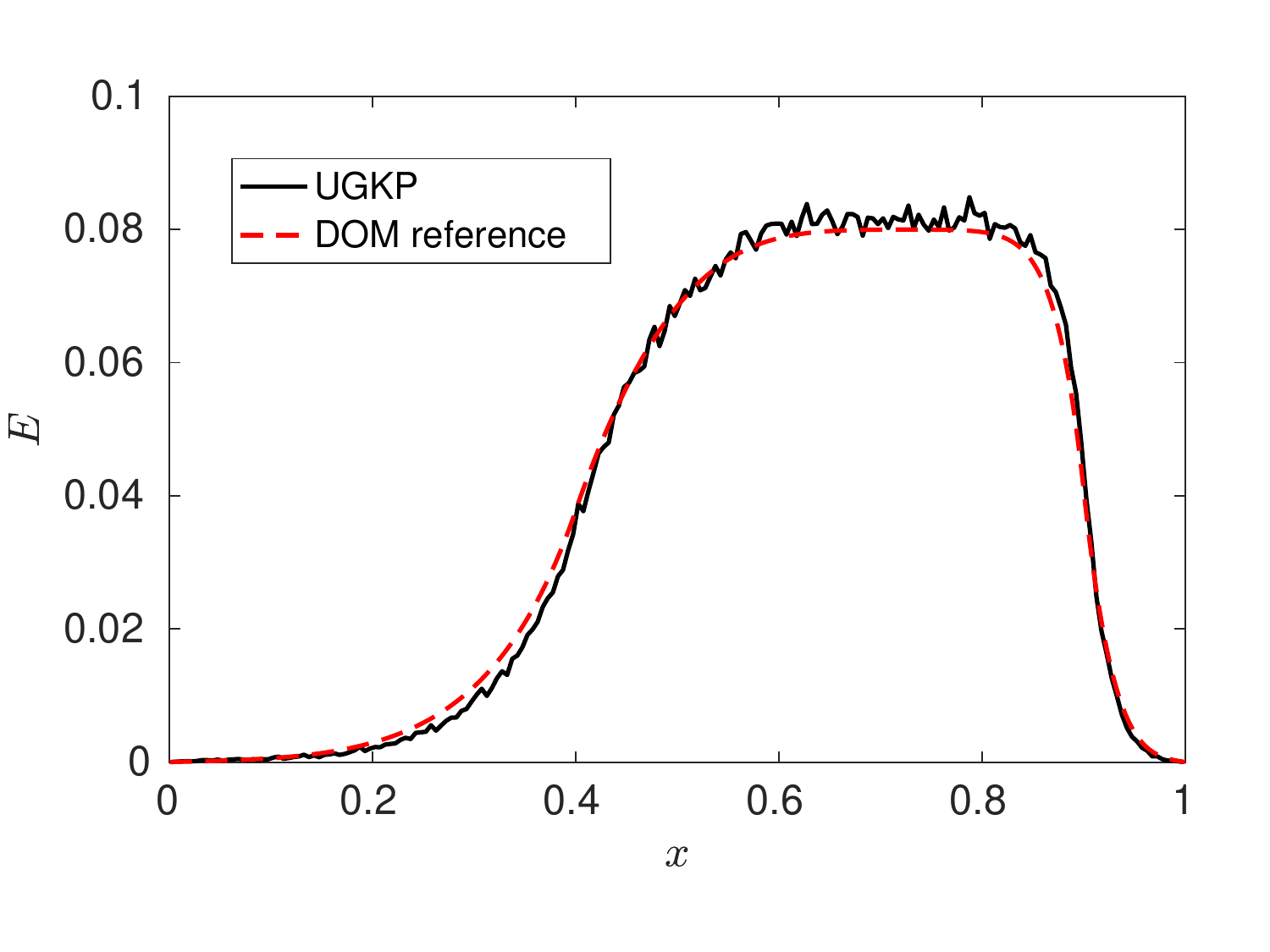}
    \caption{The macroscopic energy densities $E$ as functions of the spatial
      coordinate $x$ for the intermediate regime at $t =
      0.02$.}\label{fig:linear-intermediate}
\end{figure}

The results for $E$ are presented at time $t = 0.02$ in Figure
\ref{fig:linear-intermediate}. The reference solution is obtained from
the discrete ordinates method with $2000$ points in physical space and
$280$ points in velocity space. We observe that the result of the
UGKP method matches the reference solution
very well. It shows that the UGKP is an accurate
method in the intermediate regime.

\subsection{Radiation transfer and material energy equations}
In order to study the effectiveness of the UGKP method for radiation-material coupling, we simulated Riemann
problems for Eq. \eqref{eq:rt-coupled} in different regimes.  In
the following three examples, the parameters $\sigma$, $c$, $a$ and
$C_V$ are all set to $1$. The computation domain is $x \in
[0,1]$. The initial conditions are
\begin{equation}
  I(0, x, \mu) = \left\{\begin{array}{l}
      1, \quad \text{if}~x\in[0,\frac12), \\
        \frac12, \quad \text{if}~x \in[\frac12, 1],
      \end{array}\right.
    \end{equation}
    and $u_r = E$ for all $x$.
    Reflecting boundary conditions are
    imposed at $x = 0$ and $x = 1$. For all simulations below,
    the UGKP method always uses $200$ grids in
    space and $4000$ simulation particles within each cell. The results are obtained
    directly without employing multiple computations and averaging.

    \begin{example}[Rarefied regime]
      Take $\epsilon = 1$ and run the computation until $t = 0.3$.
    \end{example}

\begin{figure}[htbp]
      \centering
      \subfigure[Comparision of the radiation energy density $E$ between UGKP and the DOM reference
      solution.]{
        \label{fig:coupled-kinetic-E}
      \includegraphics[width=0.48\textwidth]{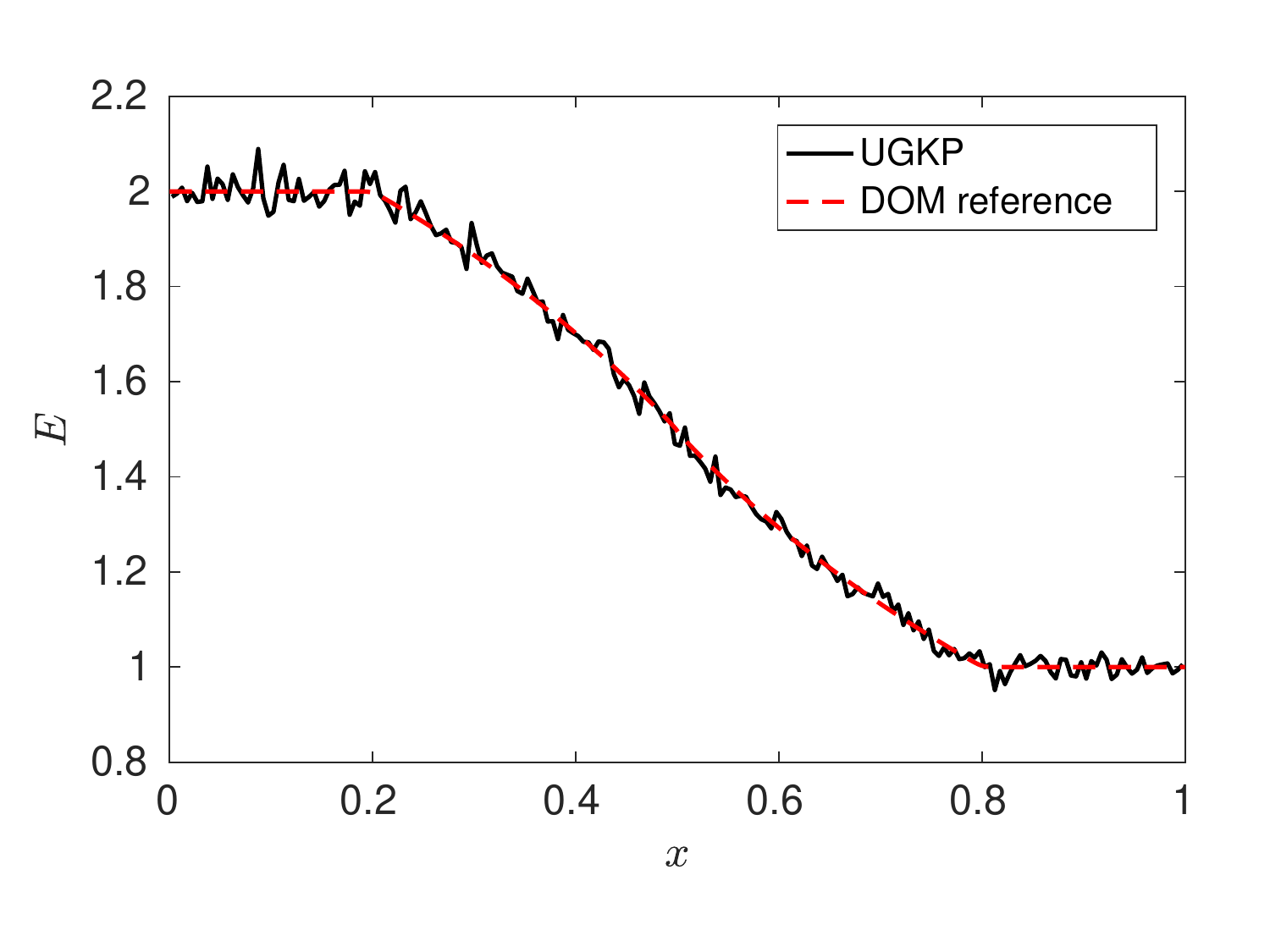}}
      \hfill
      \subfigure[Comparision between $E$ and $c u_r$ for the UGKP
      solution.]{\label{fig:coupled-kinetic-comp}
      \includegraphics[width=0.48\textwidth]{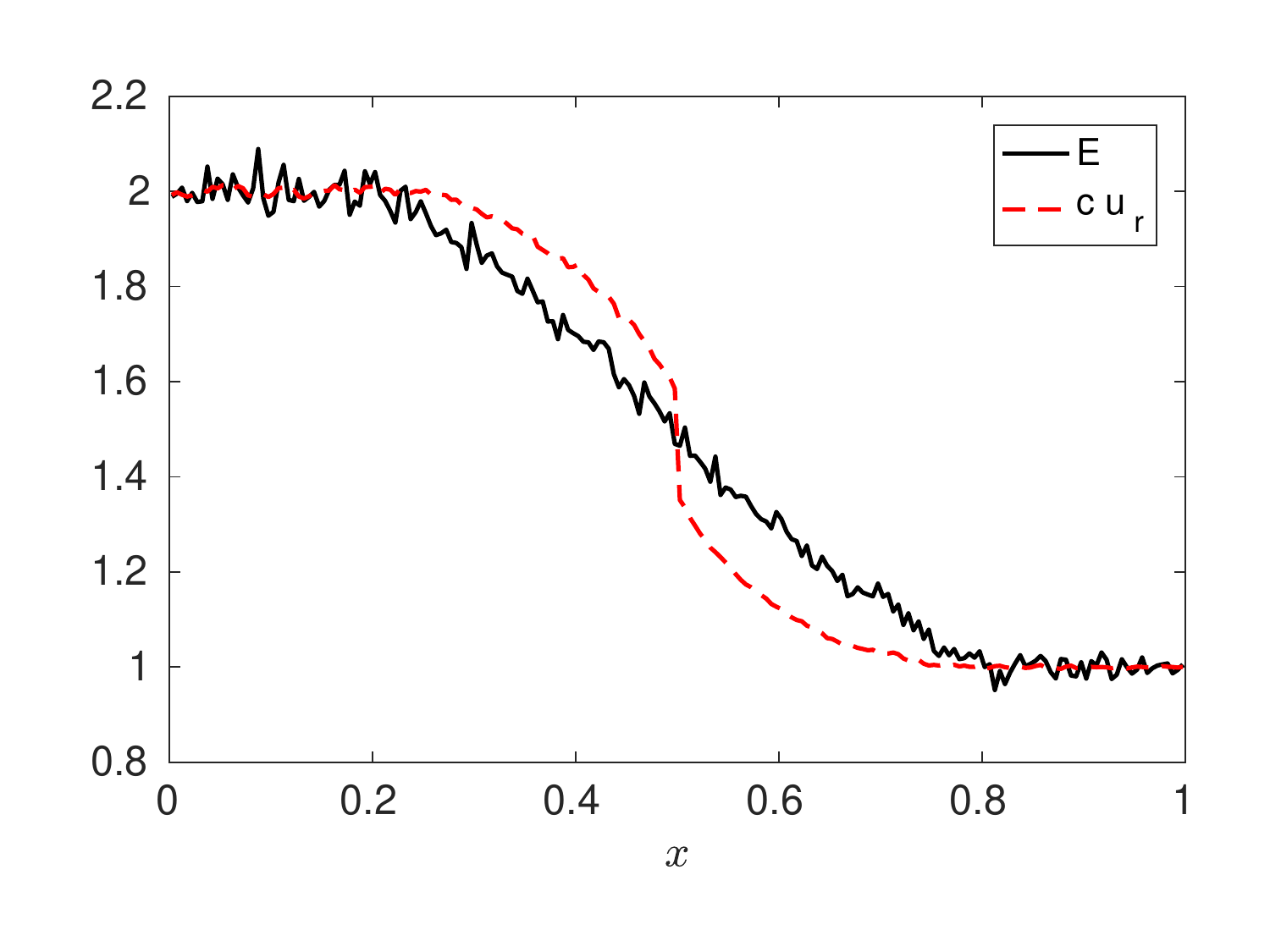}}
      \caption{Numerical results of radiation-material coupling for
      the rarefied regime at $t = 0.3$.}
      \label{fig:coupled-kinetic}
    \end{figure}

    The reference solution is obtained by employing the same splitting
    technique as outlined in Section \ref{sec:coupled}. However, for
    the reference solution, Eq. \eqref{eq:split-scattering} is
    solved with the discrete ordinates method under the finite volume
    framework, using $2000$ points in physical space and $280$ points
    in velocity space. As indicated in Fig. \ref{fig:coupled-kinetic-E},
    the solution of UGKP method is essentially the same as that of the
    reference solution. This case validates the accuracy of UGKP
    method for the problems of radiation-material
    coupling in the rarefied regime. Fig. \ref{fig:coupled-kinetic-comp} compares the UGKP solution of $E$
    and $c u_r$ and they are not in equilibrium in such a rarefied regime.

    \begin{example}[Intermediate regime]
      Take $\epsilon = 10^{-2}$ and run the computation until $t = 0.03$.
    \end{example}

\begin{figure}[htbp]
      \centering
      \subfigure[Comparision of the radiation energy density $E$ between UGKP and the DOM reference
      solution.]{
        \label{fig:coupled-intermediate-E}
      \includegraphics[width=0.48\textwidth]{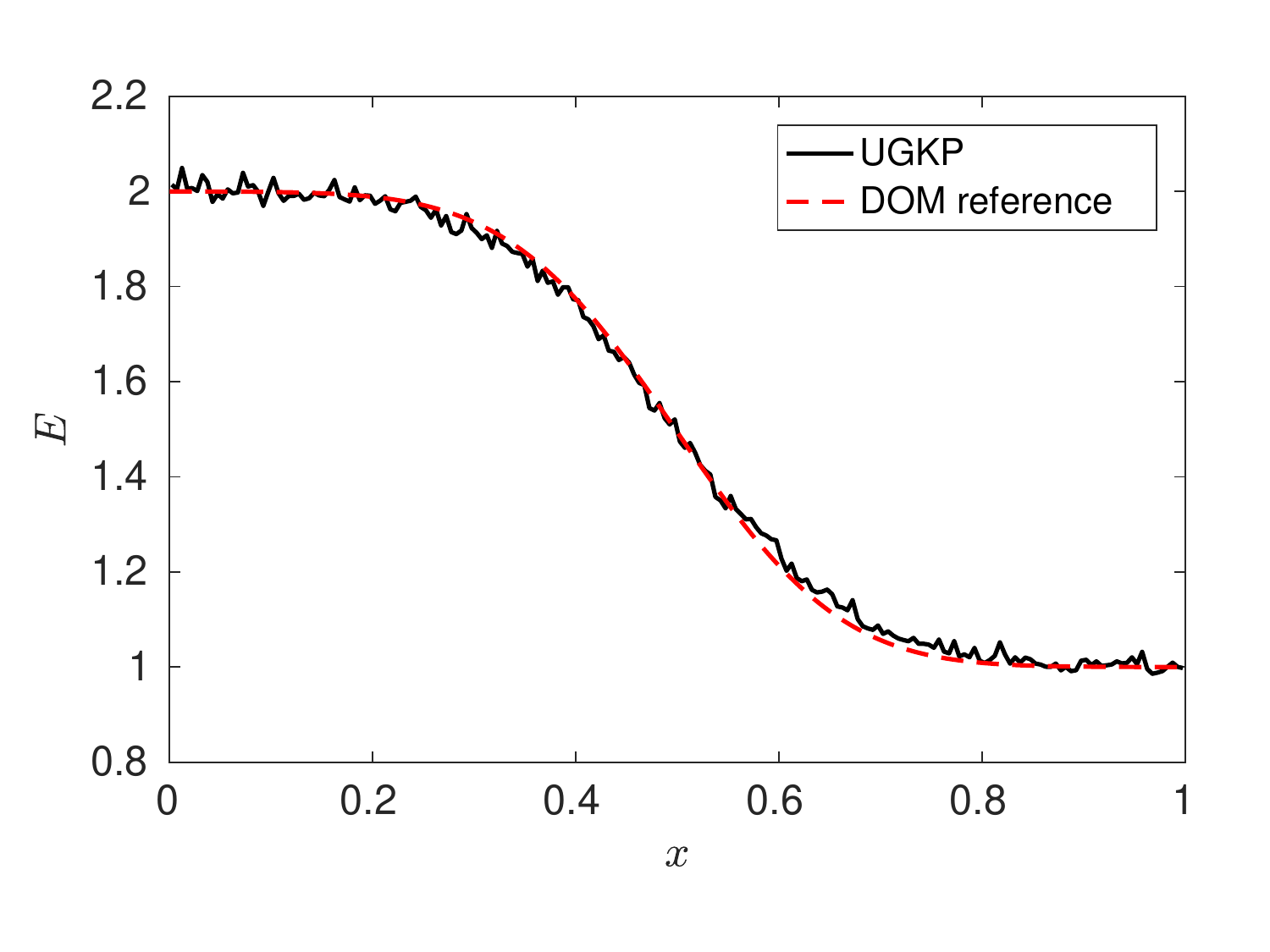}}
      \hfill
      \subfigure[Comparision between $E$ and $c u_r$ for the UGKP
      solution.]{\label{fig:coupled-intermediate-comp}
      \includegraphics[width=0.48\textwidth]{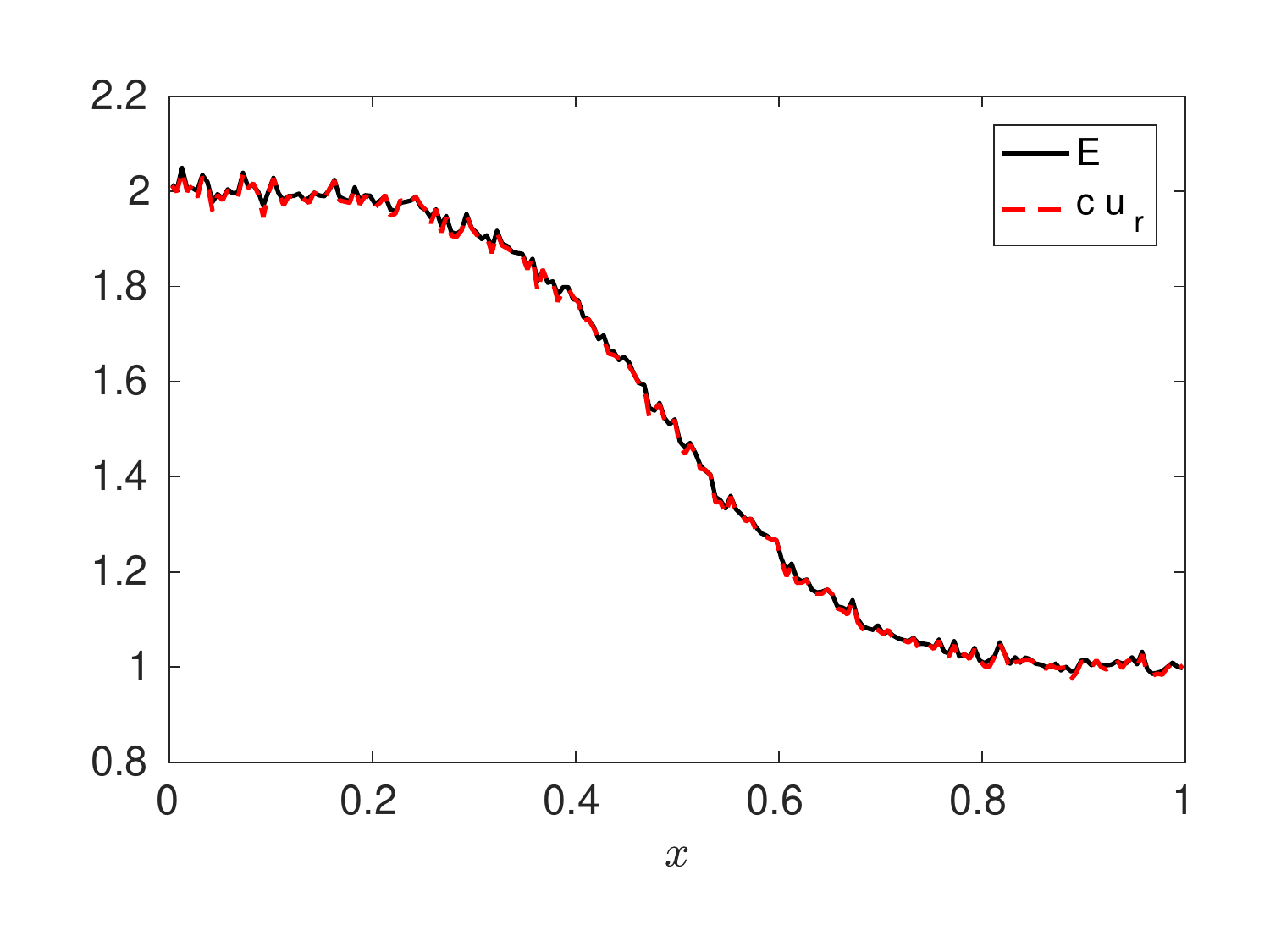}}
      \caption{Numerical results for radiation-material coupling for
      the intermediate regime at $t = 0.03$.}
      \label{fig:coupled-intermediate}
    \end{figure}

    The reference solution is obtained in the same way as in the
    previous example using the same number of discretization points.
    Fig. \ref{fig:coupled-intermediate-E} presents  the solutions of
    the UGKP method and the reference one. Both solutions are fairly consistent.
    The UGKP method is an accurate solver in the
    intermediate regime. Also, in Fig. \ref{fig:coupled-intermediate-comp} the UGKP solutions of $E$ and
    $c u_r$ are presented. In this regime, the energy exchange between
    radiation and the background medium has reached equilibrium
    at the output time.

    \begin{example}[Diffusive regime]
      Take $\epsilon = 10^{-4}$ and run the computation until $t = 0.03$.
    \end{example}

\begin{figure}%[htbp]
      \centering
      \subfigure[Comparision of the radiation energy density $E$ between UGKP and equilibrium
      diffusion solution.]{
        \label{fig:coupled-diffusive-E}
      \includegraphics[width=0.48\textwidth]{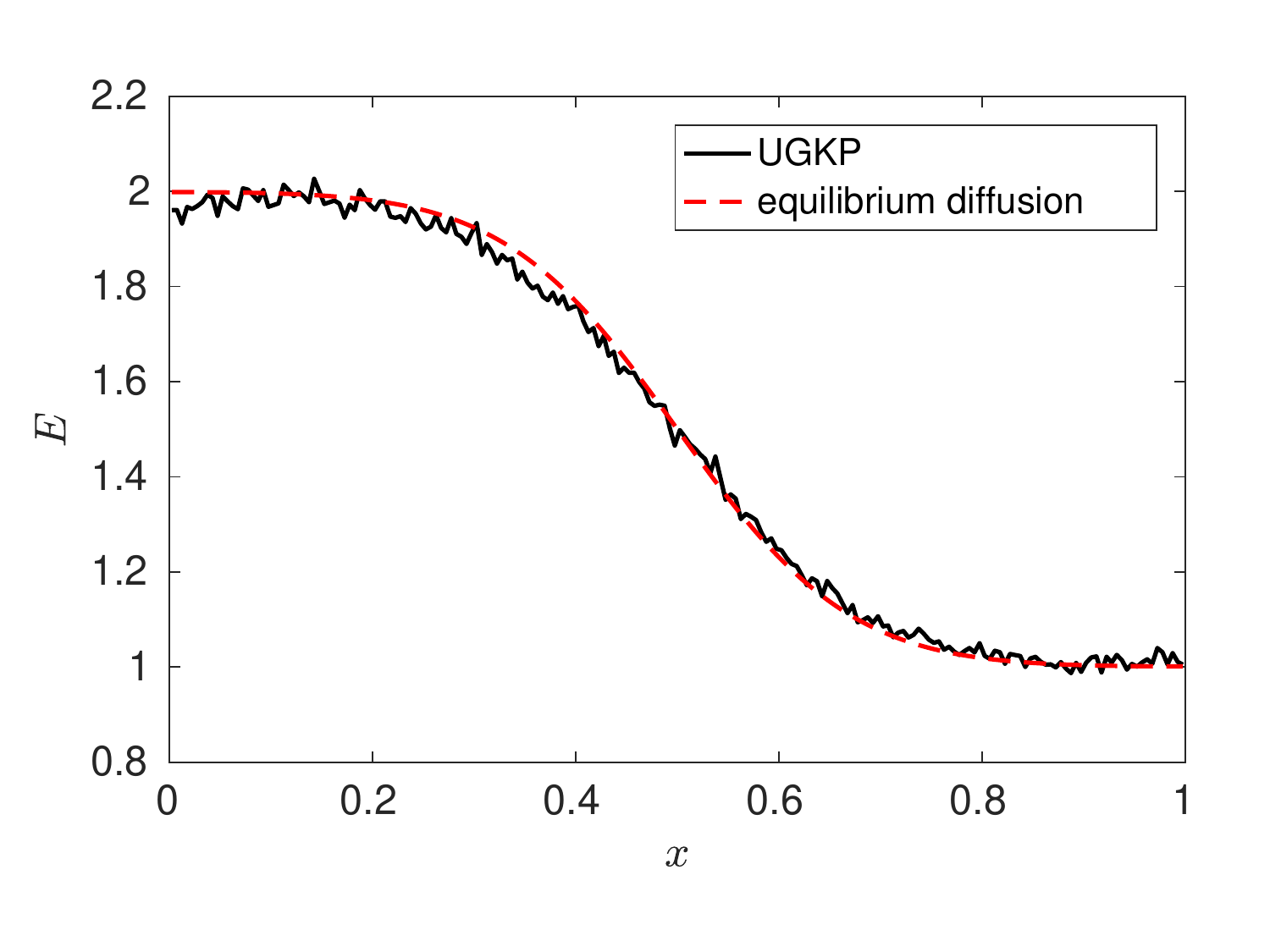}}
      \hfill
      \subfigure[Comparision between $E$ and $c u_r$ for the UGKP
      solution.]{
        \label{fig:coupled-diffusive-comp}
      \includegraphics[width=0.48\textwidth]{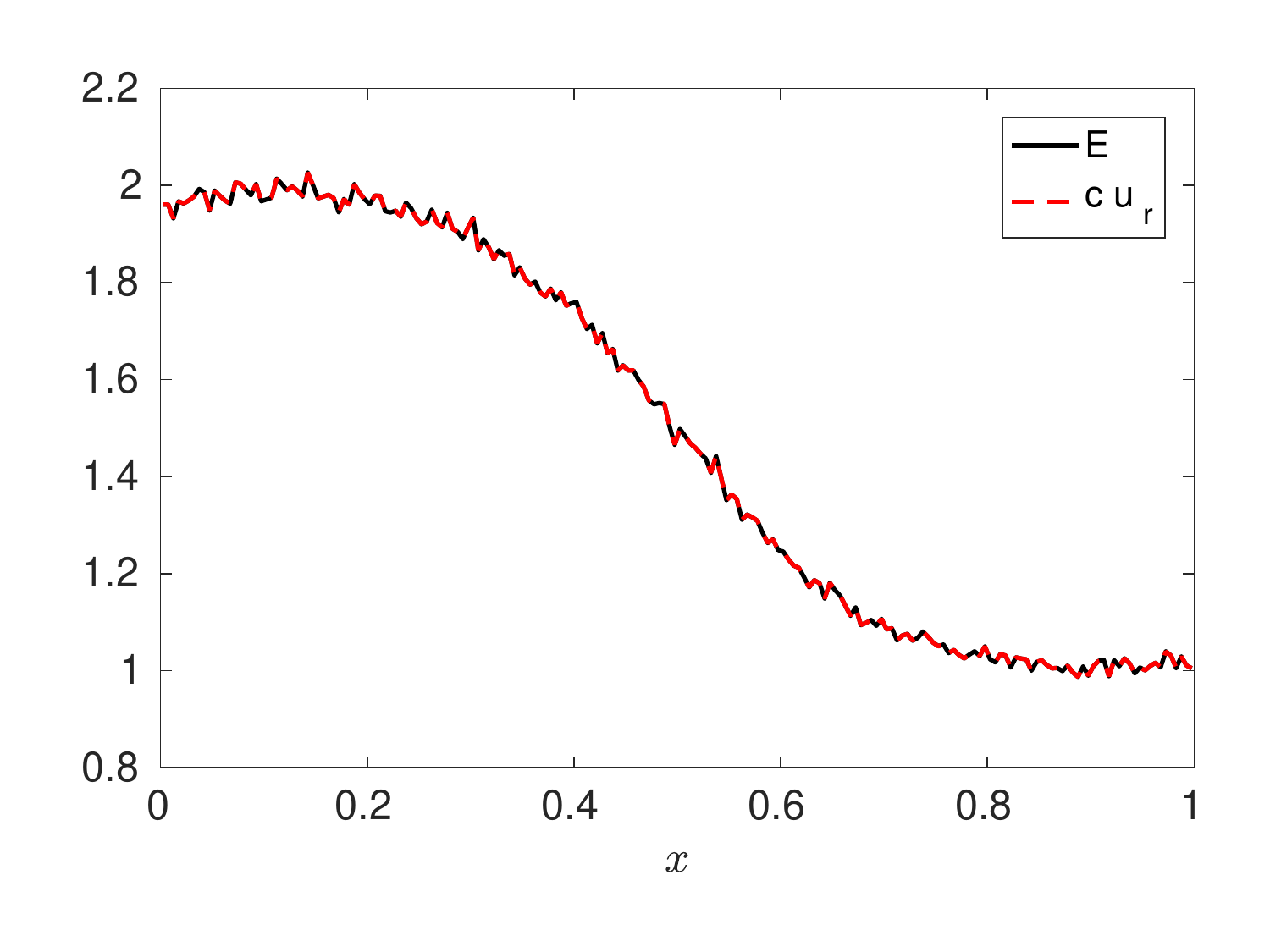}}
      \caption{Numerical results for radiation-material coupling for
      the diffusive regime at $t = 0.03$.}
      \label{fig:coupled-diffusive}
    \end{figure}

    The reference solution is obtained by solving an equilibrium
    diffusion equation with central differencing using $200$ grid points in
    space. It was observed in \cite{densmore2004asymptotic} that the
    implicit Monte Carlo method proposed by Fleck and Cummings is not
    an asymptotic preserving method for the equilibrium diffusion limit
    even though the implicit Monte Carlo method is robust and works well
    in most cases even for time steps being larger than the mean collision
    time. The UGKP solution is given in Fig. \ref{fig:coupled-diffusive-E}, which is the same as the result from the diffusion equation
    in this diffusive regime.
    Fig. \ref{fig:coupled-diffusive-comp} displays the solutions for
    radiation and material energy of UGKP
    method at $t = 0.03$, which get to equilibrium.
     This case tests  the accuracy of the UGKP method for the coupled
     radiation-material system in the diffusive regime.

\section{Conclusion}\label{sec:conclude}
In this paper, for the first time a unified gas-kinetic particle
method is proposed to simulate radiative transfer. The UGKP method is
a multiscale method for the photon transport in different regimes.
For the linear transport equation, this method recovers the solution
of the diffusion equation in the optically thick limit without
constraint on the time step being less than the photon's mean
collision time. At the same time, it gives the exact solution in the free
transport regime.  The UGKP method is also extended to the coupled
radiation-material system. With the inclusion of energy exchange, the
UGKP method can give excellent simulation results in different
regimes.  A few benchmark problems are tested to show the performance of the current scheme.
The accuracy and efficiency of the UGKP method are fully
confirmed.  In the future work, we will extend this method to
multidimensional and frequency-dependent radiative transfer problems.

%\section*{References}
%\newpage
\bibliography{references}

\end{document}